\documentclass[12pt]{article}
\usepackage{graphicx}
\usepackage{setspace}
\usepackage{amssymb}
\usepackage[left=1in,right=1in,top=1in,bottom=1in]{geometry}
\def \be {\begin{equation}}
\def \ee {\end{equation}}
\def \ba {\begin{array}}
\def \ea {\end{array}}
\def \bea {\begin{eqnarray}}
\def \eea {\end{eqnarray}}

\def \ble {\begin{widetext}\begin{equation}}
\def \ele {\end{equation}\end{widetext}}
\def \blea {\begin{widetext}\begin{eqnarray}}
\def \elea {\end{eqnarray}\end{widetext}}

\def \blea {\begin{widetext}\begin{eqnarray}}
\def \elea {\end{eqnarray}\end{widetext}}

\begin{document}
\begin{titlepage}
\hfill

\vspace*{20mm}
\begin{center}
{\Large \bf Measuring the Black Hole Interior from the Exterior}

\vspace*{15mm}
\vspace*{1mm}

Wu-zhong Guo

\vspace*{1cm}

{School of Physics, Huazhong University of Science and Technology,\\
Luoyu Road 1037, Wuhan, Hubei 430074, China\\
{\it wuzhong@hust.edu.cn}
}

\vspace*{1cm}
\end{center}

\begin{abstract}
In this essay, we argue that an observer outside the horizon can reconstruct the geometry of a black hole's interior through external measurements. This procedure builds on recent studies of the holographic duality of timelike entanglement entropy and its connection to spacelike entanglement entropy. Furthermore, we propose that this phenomenon reveals a fundamental correlation between the degrees of freedom inside and outside the black hole at the level of classical spacetime.
\end{abstract}
\vskip 2cm

\begin{center}
Essay written for the Gravity Research Foundation \\ 2025 Awards for Essays on Gravitation
\vskip 1cm

March 30, 2025
\end{center}
\end{titlepage}

\textbf{Introduction}

Black holes might be the most fascinating objects in the sky, even though we can't directly ``see'' them. Partial information about a black hole is hidden behind the horizon, but fortunately, when quantum effects are considered, Hawking radiation can carry some of it to outside observers. This suggests that the black hole interior is connected to the external degrees of freedom. Understanding the correlation between the inside and outside of the horizon is crucial to the information paradox, see the recent review \cite{Almheiri:2020cfm} and references therein for the latest developments. A question arises: can we observe this correlation using only classical spacetime? The answer is yes, and this is precisely what we aim to demonstrate in this essay.

In classical spacetime, all information is encoded in the metric $g_{\mu\nu}$. For a static black hole spacetime, a horizon exists, dividing the spacetime into two regions. An observer, Alice, who remains outside the horizon, can only detect and perform measurements in the exterior region, while she cannot directly observe Bob, who is inside the horizon. In quantum mechanics, their correlation naturally arises through non-vanishing correlation functions in the black hole background. Moreover, the idea of black hole complementarity suggests that the black hole interior can be reconstructed from the exterior degrees of freedom \cite{Stephens:1993an,Susskind:1993if}.

At the classical level, this picture breaks down, as measurements are tied to the spacetime metric, such as the length of an object nearby. In this essay, we aim to demonstrate that Alice, while remaining outside the horizon, can determine certain geometric objects inside the horizon solely through measurements made in the exterior region. Furthermore, we argue that this can be interpreted as evidence of correlation even in classical spacetimes.

\textbf{Timelike and spacelike holographic entanglement entropy}

\begin{figure}[htbp]
    \centering
    \includegraphics[width=0.8\textwidth]{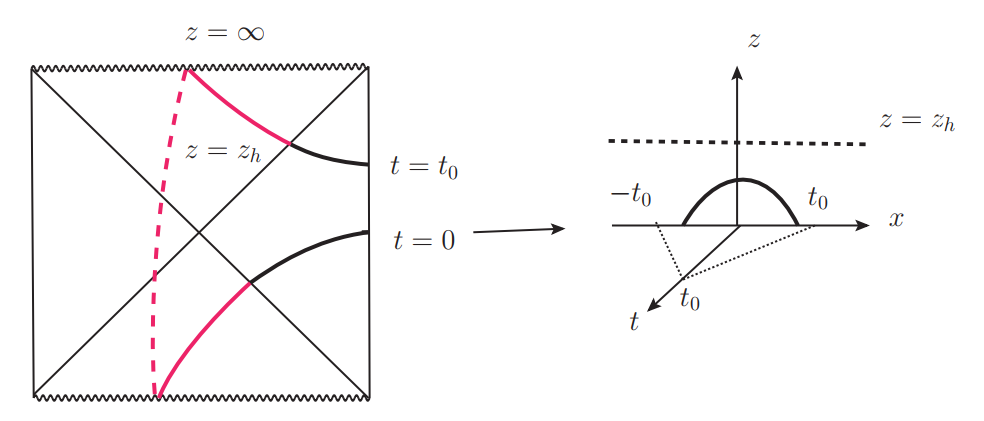}
    \caption{The RT surface for the timelike strip is divided by the black hole horizon into two distinct regions. Inside the horizon, the RT surface is depicted by the red solid and dashed lines, while the black solid line extending from the horizon to the boundary represents the portion outside. The relation (\ref{sumrule}) implies that the area of the RT surface inside the horizon can be expressed as the sum of the exterior RT surface area and additional contributions from RT surfaces associated with spacelike subregions (black bold solid line) on the Cauchy surface $t=0$ (as shown in the right panel). }
    \label{sumrule_pic}
\end{figure}

Our argument relies on holographic Ryu-Takayanagi surfaces associated with both timelike and spacelike entanglement entropy (EE) \cite{Ryu:2006bv}. To set the stage for our discussion, we first introduce the necessary concepts and computations. Specifically, we consider a black hole in an asymptotically AdS$_{d+1}$ metric,
\bea\label{set_up_metric}
ds^2=\frac{1}{z^2}\left(-f(z)dt^2+\frac{dz^2}{f(z)}+dx^2+d\vec{y}^2 \right),
\eea
where the AdS radius $L=1$ and $d\vec{y}^2=\sum_{i=1}^{d-2}dy_i^2$. The AdS boundary is at $z = 0$.  We assume there exists a Killing horizon $z=z_h$ with $f(z_h)=0$  for the Killing vector $\partial_t$. The Penrose diagram of the extended eternal black hole is shown in Fig. \ref{sumrule_pic}.
In the context of AdS/CFT, this spacetime corresponds to the thermofield double state of two entangled CFTs residing on the two asymptotic boundaries \cite{Maldacena:2001kr}. Information inside the black hole horizon can be accessed through certain observables in the dual CFTs, such as correlators of heavy operators located at the two boundaries or the holographic EE of spacelike subsystems living on both boundaries \cite{Fidkowski:2003nf,Hartman:2013qma}. The dual bulk geometric objects associated with these observables can extend into the black hole interior.

\begin{figure}[htbp]
    \centering
    \includegraphics[width=0.6\textwidth]{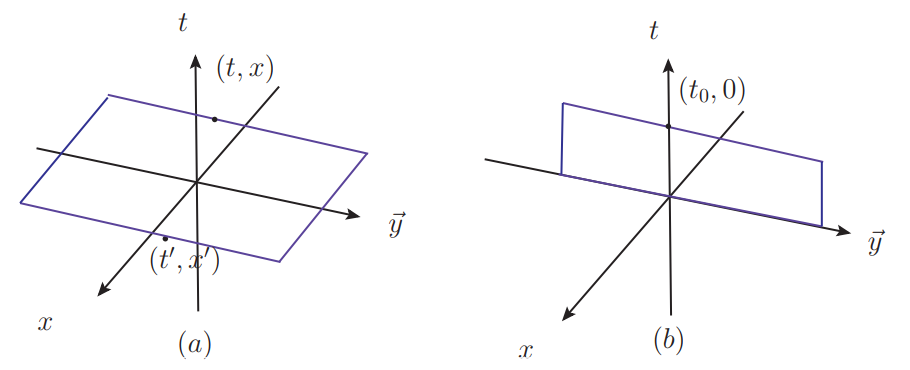}
    \caption{Illustration of the strip we consider. (a) represents the general case, where the subregion can be timelike or spacelike depending on the sign of  $\Delta s^2$. (b) depicts a special case of a timelike strip.}
    \label{strip_pic}
\end{figure}

Here, we consider the holographic entanglement entropy for timelike subregions on one boundary. Specifically, we focus on a general strip subregion, denoted by $s(t,x;t',x')$, as shown in Fig.\ref{strip_pic}. We define $\Delta s^2=-(t-t')^2+(x-x')^2$. The subregion is timelike for $\Delta s^2<0 $, while it is spacelike for $\Delta s^2>0$. Previous studies have primarily focused on entanglement entropy for spacelike subregions, and the physical meaning of timelike EE remains unclear. However, in QFTs, timelike EE can be well-defined through the analytical continuation of twist operators. In general, the results are complex-valued \cite{Doi:2022iyj}. While the timelike EE is well-defined and computable in QFTs, its holographic dual remains a subject of debate. Nonetheless, we believe that the Ryu-Takayanagi formula $S=\frac{\mathcal{A}}{4G}$ is still applicable, where $\mathcal{A}$ denotes the area of the RT surface in the bulk. 

Let us briefly summarize some significant aspects of timelike and spacelike EE.
\begin{itemize}
\item In QFTs, timelike EE can be defined through the analytical continuation of correlation functions involving twist operators.
\item Two proposals for holographic timelike EE:

(i) The RT surface should include both the spacelike surface and its timelike counterpart, with the timelike surface contributing to the imaginary part of the timelike EE \cite{Doi:2022iyj}.

(ii) The RT surfaces are given by extremal surfaces in a complexified geometry \cite{Heller:2024whi}.
\item In the AdS$_3$ case, both proposals yield the same results. However, distinctions arise in higher-dimensional examples, such as the timelike strip.
\item Both proposals suggest that the RT surface extends into the horizon, thus detecting the interior of the black hole.
\item The black hole horizon acts as a barrier for the RT surface of spacelike subregions, meaning that holographic EE for spacelike subregions cannot access information inside the black hole \cite{Engelhardt:2013tra}.
\end{itemize}

Interestingly, timelike EE can be related to spacelike EE in an exact manner. In the following, we will explain why such a relationship is expected to exist. In QFTs, the standard method for computing EE is via replica methods. The evaluation of EE can be translated into correlation functions involving twist operators. In 2-dimensional QFTs, the twist operators are local operators, whereas in higher dimensions, the twist operators are non-local and located at the boundaries of subregions. For the timelike strip $s(0,0;t_0,0)$, as shown in Fig.\ref{strip_pic}, the timelike EE is related to the correlator $\langle \Sigma_n (t_0,0)\Sigma_n(0,0)\rangle_{QFT^n}$, where $\Sigma_n$ denotes the operators in the $n$-copied theories located at the two boundaries of the strip. This correlator can be evaluated through an analytical continuation of the Euclidean correlator via the standard Wick rotation $\tau \to i t$. Following causality, the operator $\Sigma_n(t_0,0)$ can be decomposed as a combination of operators in the spacelike strip $A_s:=s(0,-t_0;0,t_0)$ on the Cauchy surface $t=0$, as shown in Fig.\ref{sumrule_pic}. Formally, we have
\bea\label{operators_relation}
\Sigma_n(t_0,0)=\sum_{k=0}^{\infty}\int_{A_s}d^{d-1}xf_k(t_0,\vec{x})\partial_k\Sigma_n(0,\vec{x})+..., 
\eea
where ``...'' denotes contributions from other possible operators, and $f_k(\vec{x})$ are functions supported in region $A_s$. As a result, the two point correlator $\langle \Sigma_n (t_0,0)\Sigma_n(0,0)\rangle_{QFT^n}$  can be associated with correlators involving operators located in the spacelike subregion $A_s$. Unfortunately, (\ref{operators_relation}) is only a formal expression. We do not expect that one could solve for $f_k$ and the ``...'' terms, even for free field theories. However, (\ref{operators_relation}) provides a guideline that there may exist a relationship between timelike EE and spacelike correlations, which offers a way to understand the physical meaning of timelike entanglement.

For the thermal states dual to the black hole described by (\ref{set_up_metric}), we can directly establish a relationship between timelike and spacelike EE. Consider $d=2$ as an example, where the subregion $s(t,x;t',x')$  corresponds to intervals. Denote the EE for the interval $s(t,x;t',x')$ by $S(t,x;t',x')$. We can derive the following relationship \cite{Guo:2024lrr}
\bea\label{sumrule}
S(0,0;t_0,0)=\frac{1}{2}\int_{-t_0}^{t_0}dx f_0(t_0,x)S(0,x;0,0) +\frac{1}{2}\int_{-t_0}^{t_0}dxf_1(t_0,x)\partial_t S(0,x;0,0),
\eea
where $f_0(t_0,x)=\delta(x+t_0)+\delta(x-t_0)$ and $f_1(t_0,x)=1$. The timelike EE is associated with the spacelike EE and its first-order temporal derivative. It is also possible to extend these results to higher-dimensional black holes, although analytical results for both timelike and spacelike EE are not obtainable. If $t_0\ll z_h$, we can perturbatively evaluate the results. We can perturbatively show that a similar relationship to (\ref{sumrule}) holds in the higher-dimensional case \cite{Guo:2024lrr}.

\textbf{Measuring the interior geometry from outside observers}

Now, let us explore the holographic interpretation of the relation (\ref{sumrule}). The key insight is that the RT surfaces (or geodesic lines in AdS$_3$) for timelike EE can extend into the black hole's interior, whereas those for spacelike EE remain confined outside the horizon. The RT surface associated with timelike EE should be anchored on the boundary subregion $s(0,0;t_0,0)$, traverse the horizon, and extend toward the singularity, as illustrated in Fig. \ref{sumrule_pic}. Additionally, we must account for extremal surfaces connecting the singularities, represented by the dashed red lines in Fig. \ref{sumrule_pic}. 

In AdS$_3$ case, as we have mentioned two different proposals exist for choosing the geodesic lines that connect the singularities. In \cite{Doi:2022iyj}, the authors suggest that the connection is given by a timelike geodesic, which can be determined via the stationarity condition. In contrast, \cite{Heller:2024whi} proposes an analytic continuation of the geodesic into a complexified geometry. Furthermore, in \cite{Guo:2024lrr}, a method is introduced to fix the geodesic by analytically continuing the Euclidean result, which can be generalized to higher dimensions. Nevertheless, the contribution from these geodesics connecting the singularities is a constant term $i\pi$, which corresponds to the imaginary part of the timelike EE. The real part, on the other hand, is determined by the spacelike geodesic lines that connect the boundary to the singularity.

 The horzion at $z=z_h$ divides the geodesic into two segments, so the timelike EE naturally decomposes into contributions from RT surfaces inside and outside the horizon. Meanwhile, the right-hand side of (\ref{sumrule}) corresponds to the EE of spacelike subregions in the region $x\in [-t_0,t_0]$ on the Cauchy surface $t=0$, as shown in the right panel of Fig.\ref{sumrule_pic}. Therefore, the relation (\ref{sumrule}) establishes a connection between the lengths of RT surfaces inside and outside the horizon. This provides a practical method for probing the geometry of the black hole interior using only exterior information.

For an observer like Alice, who remains outside the horizon, directly accessing the area of RT surfaces inside the black hole is impossible. However, she can measure the area of RT surfaces that connect the boundary to the horizon, as well as those associated with spacelike subregions on the Cauchy surface—represented by the bold black solid lines in Fig. \ref{sumrule_pic}. By gathering this information and applying the relation (\ref{sumrule}), she can infer the area of the extremal surfaces inside the horizon, depicted by the red lines in Fig. \ref{sumrule_pic}.
By varying the strip width $t_0$, the corresponding RT surfaces sweep through both the exterior and interior regions of the black hole. In principle, with sufficient measurements, Alice could reconstruct the black hole's interior geometry and extract information about its internal structure—despite never directly receiving signals from within the horizon.

Moreover, the above process also provides us with a way to understand the correlation between the interior and exterior degrees of freedom of the black hole at the level of classical spacetime, then offers an intrinsic perspective on the nature of gravity. The argument of black hole complementarity suggests that the degrees of freedom inside the black hole can be reconstructed from outside information. This highlights the crucial role that entanglement plays in understanding complementarity.

In QFTs, the entanglement between a subregion and its complement is generally divergent. One consequence of this is that it is possible to construct a set of states that is dense in the Hilbert space using \textit{only} operations within the subregion. This is the well-known Reeh-Schlieder theorem in algebraic QFTs. In gravity theory, it is also believed that entanglement is crucial for the classical connectivity of spacetime \cite{VanRaamsdonk:2010pw}. However, connectivity is a classical concept of spacetime. If a similar Reeh-Schlieder theorem exists for gravity, it would imply that the metrics for a subregion and its complementary part are related in some way.

Here, the spacetime inside and outside the black hole is clearly connected and thus entangled. Therefore, the relationship (\ref{sumrule}) reflects the correlation of metrics for different subregions in gravity theory at the classical level. Quantum entanglement is a key feature in quantum mechanics, but here we have observed the correlation between the interior and exterior geometry even without considering quantum effects.

For the BTZ black hole, we can construct the exact relation (\ref{sumrule}). But is this process valid for more general cases? In higher-dimensional black holes, we have only constructed the relation through perturbative calculations so far. However, we can argue that this phenomenon should hold for more complicated scenarios, such as when matter or gravitational waves are added to the spacetime, or even beyond AdS spacetime. In QFTs, we expect that timelike EE, which can be computed using correlation functions with timelike separation, should be related to spacelike correlation functions via (\ref{operators_relation}). Eq.(\ref{operators_relation}) follows from causality and is an operator relation, meaning the argument is independent of specific states.

In more general cases, we should also consider possible contributions from higher temporal derivatives of spacelike EE or correlation functions of other operators. These operators are only located in the spacelike subregions $A_s$. Therefore, according to the subregion/subregion duality in AdS/CFT \cite{Czech:2012bh}, the correlation functions can be constructed using the bulk entanglement wedge for the spacelike region $A_s$. The entanglement wedge is associated with the regions surrounded by RT surfaces, which lie outside the horizon. These correlation functions are inherently tied to information accessible outside the horizon.

In summary, we expect that the timelike EE for the timelike strip $s(0,0;t_0,0)$ can be reconstructed from the bulk quantities in the entanglement wedge for the spacelike strip $s(0,-t_0;0,t_0)$. The horizon divides the RT surfaces for timelike EE into two parts. Therefore, the area of the interior portion of the RT surface can be reconstructed using information from outside the horizon. This suggests that the geometry of the black hole interior can be determined solely by measuring quantities outside the horizon.

The singularity hidden within the horizon is the primary feature we aim to detect. Let's consider another important question: can the procedure we discussed earlier allow us to learn about the singularity from measurements made by outside observers? It's important to note that the area of the portion of the RT surface inside the horizon remains finite, even though it extends all the way to the singularity. This is expected, as an observer falling into the black hole will eventually reach the singularity after a finite amount of proper time. Similarly, for the spacelike extreme surface, the situation is analogous. However, an observer falling toward the singularity would perceive its existence by noticing the increasingly intense tidal forces they experience as they approach it. In this context, we can consider two strip subregions: $s(0,0;t_0,0)$ and $s(0,0; t_0+\delta t_0,0)$. The areas of the RT surfaces for these two strips will show a difference near the singularity. By examining this difference, we can extract information about the effect of the singularity.

As we have argued, the interior geometry can be reconstructed from measurements made outside the black hole. Therefore, it is possible to detect the signature of the singularity based solely on external measurements. Here, we have provided a general argument, but it is also possible to design a thought experiment based on this idea to further explore this concept. 

\textbf{Summary and discussion}

In this essay, we demonstrate that it is possible to determine the geometry of a black hole's interior through measurements made by outside observers. This can be viewed as evidence of a correlation between the interior and exterior of the black hole, even without considering quantum effects. Perhaps, for a real black hole in our universe, a similar procedure could allow us to ``see'' its interior through external measurements.\\

\textbf{Acknowledgements}

I am supposed by the National Natural Science Foundation of China under Grant No.12005070 and the Hubei Provincial Natural Science Foundation of China
under Grant No.2025AFB557.

\end{document}